\documentclass[%
 reprint,
superscriptaddress,
 amsmath,amssymb,
 aps,
]{revtex4-2}

\usepackage{graphicx}
\usepackage{dcolumn}
\usepackage{bm}
\usepackage{hyperref}

\usepackage{color}

\begin{document}
\preprint{APS/123-QED}

\title{Drying-induced film formation in redispersible polymer powder: \\a time-resolved mechanical study}

\author{Justine Henry}
 \affiliation{ENSL, CNRS, Laboratoire de Physique, F-69342 Lyon, France}
 \affiliation{Saint-Gobain Recherche Paris, 93303 Aubervilliers, France}
 \author{Sébastien Manneville}
  \affiliation{ENSL, CNRS, Laboratoire de Physique, F-69342 Lyon, France}
  \affiliation{Institut Universitaire de France (IUF)}
 \author{Ludovic Feige}
\affiliation{Saint-Gobain Recherche Paris, 93303 Aubervilliers, France}
 \author{Clara Paillard}
\affiliation{Saint-Gobain Recherche Paris, 93303 Aubervilliers, France}
\author{Thibaut Divoux}
\email{Thibaut.Divoux@ens-lyon.fr}
 \affiliation{ENSL, CNRS, Laboratoire de Physique, F-69342 Lyon, France}

\date{\today}

\begin{abstract}
The mechanical evolution of latex dispersions during drying remains poorly understood, despite its importance for coatings, adhesives, and construction materials. 
Here, we use time-resolved mechanical spectroscopy to follow the formation of a polymer film from a dense aqueous latex suspension. The viscoelastic spectra retain a power-law frequency dependence and collapse onto a master curve using time-dependent shift factors, revealing a time-aging superposition principle. 
Moreover, the frequency shift factor eventually plateaus, suggesting a crossover toward arrested particle dynamics and film formation. 
This scenario is robustly observed at various temperatures, allowing us to determine an apparent activation energy associated with film formation.
Finally, the non-linear response of the dispersion, measured throughout film formation, shows a ductile-to-brittle transition, consistent with the emergence of a solid-like film. Together, these results establish a rheological picture of the transformation from a dense colloidal suspension into a continuous polymer film. 
\end{abstract}

\maketitle

\textit{Introduction.-} Latex dispersions consist of colloidal polymer particles suspended in a volatile solvent and are widely used in coatings, adhesives, inks, and construction materials. Their technological importance stems from their ability to transition from a liquid-like dispersion into a continuous polymer film upon drying, thereby providing cohesion and mechanical strength to the final material \cite{Keddie:2010,Steward:2000}.
Film formation results from the progressive removal of solvent, which increases the particle volume fraction until a densely packed assembly is reached. Capillary forces then promote particle deformation and the elimination of interstitial voids, while polymer chains diffuse across particle boundaries, ultimately leading to particle coalescence and the formation of a mechanically coherent film \cite{Joanicot:1996,Keddie:1997,Winnik:1997}. 

Most experimental studies have characterized film formation through the structural evolution of drying dispersions. Optical, scattering, gravimetric, atomic force microscopy, and NMR techniques have provided valuable information on solvent transport, particle packing, and coalescence dynamics \cite{Chevalier:1992,Butt:1994,Lin:1995,Voogt:2019,Arjmandi:2024}. 
In comparison, the evolution of the mechanical response during drying has received much less attention. For latex dispersions, Molenaar \textit{et al.} \cite{Molenaar:1997} monitored the apparent viscosity of drying latex dispersions using a rolling-ball technique, while substrate bending has been used to characterize the development of internal stresses during film formation \cite{Tirumkudulu:2004,VonDerEhe:2007,Yow:2010}. 
Other approaches, developed for colloidal dispersions more generally, include measurements of drying-induced normal stresses \cite{Bouchaudy:2019} and time-resolved rheometry under controlled normal force \cite{Lehericey:2021}. Lin and Meier used atomic force microscopy to follow the kinetics of film formation, and established that this kinetic obeys a time-temperature superposition  \cite{Lin:1996}. However, the evolution of the frequency-dependent viscoelastic response throughout latex film formation remains largely unexplored.

In this Letter, we introduce an original approach to monitor the rheological evolution of a dense redispersible polymer powder suspension throughout film formation. By exploiting the growth of a peripheral coalesced latex film surrounding a liquid core of latex particles in a parallel-plate geometry, we extract the viscoelastic contribution of the forming film from time-resolved mechanical spectroscopy (TRMS). The resulting spectra display a robust power-law frequency dependence throughout drying and collapse onto a master curve through time-dependent shift factors, revealing a time-aging superposition principle during film formation. In addition, oscillatory shear experiments of increasing amplitude conducted at various drying times reveal a ductile-to-brittle transition that accompanies the progressive growth of a mechanically coherent film. These results shed new light on the rheological pathway from a dense latex dispersion to a mechanically coherent polymer film.

\textit{Materials and methods.-} Experiments are performed at $T=25^\circ \rm C$ (unless stated otherwise) on a dense aqueous suspension ($\phi=62.5~\%~\rm vol$) of industrial latex particles prepared from a redispersible polymer powder (RDP, Saint-Gobain) of an ethylene-vinyl acetate (EVA) copolymer ($T_g = 18 ^\circ$C). The commercial powder consists of grains (of size $100~\mu\rm m$) containing latex particles coated with a mixed shell of hydrophilic  Poly(vinyl alcohol) (PVOH) and hydrophobic EVA polymers (about $30~\rm nm$ thick) \cite{Wu:1996,Borget:2005}, together with kaolin platelets and calcium carbonate crystals (see SEM images in Fig.~S1 in the Supplemental Materials, SM). Upon dispersion in deionized water, the grains dissociate, releasing the latex particles, which are subsequently homogenized under high shear (1000~rpm during $180~\rm s$) using a mechanical stirrer (IKA RW 20 Digital mixer equipped with a four-bladed propeller). The resulting suspension displays a broad but predominantly bimodal particle size distribution centered around $1~\mu \rm m$ and $10~\mu \rm m$ (see Fig.~S2 in the SM). 

 \begin{figure}[t!]
    \centering
    \includegraphics[width=0.9\linewidth]{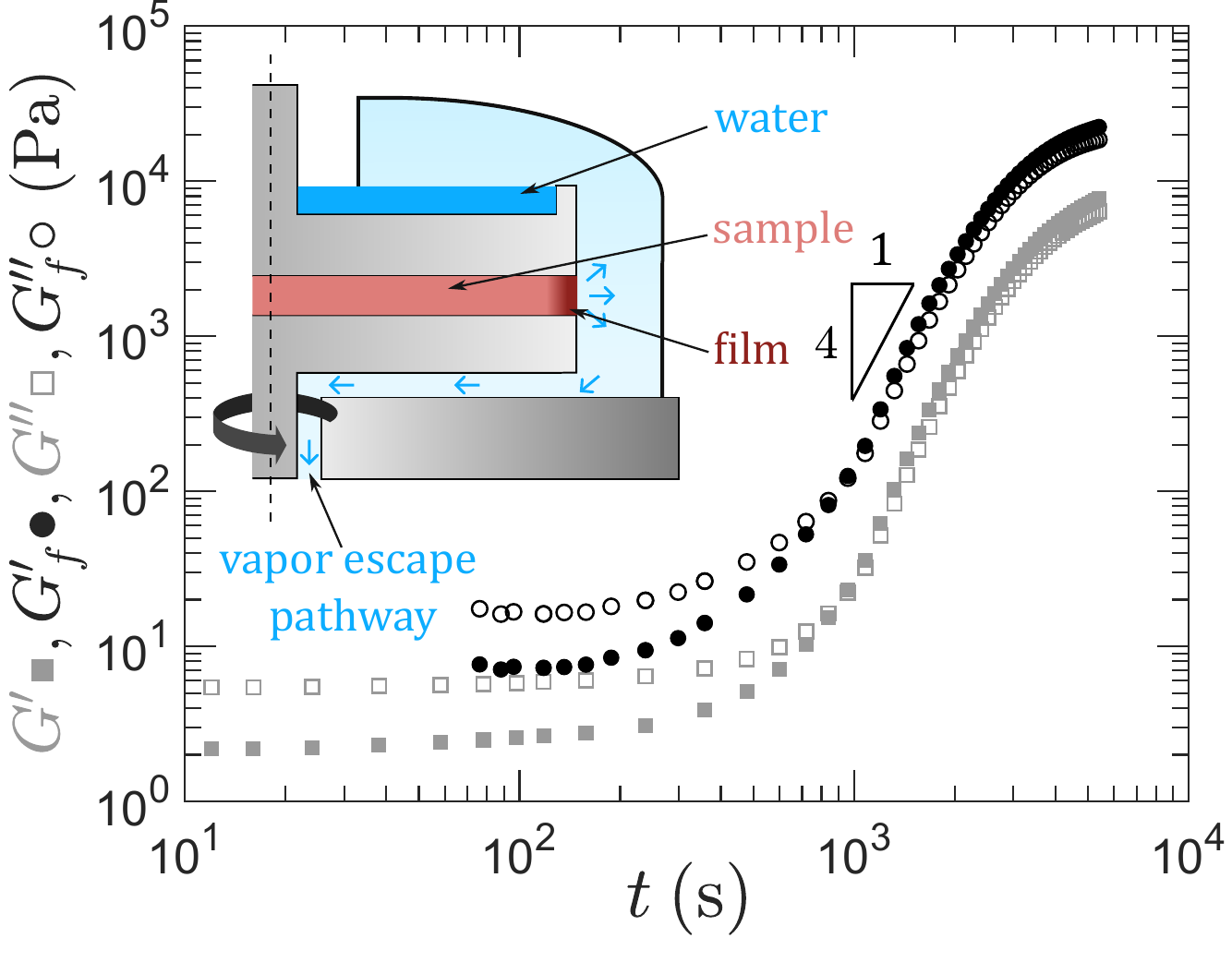}
    \caption{Viscoelastic moduli of a dense latex suspension ($\phi=62.5~\%~\rm vol$) measured after preshear under small-amplitude oscillatory shear ($\omega=2\pi~\rm rad.s^{-1}$, $\gamma_0=1\%$). Gray symbols: raw rheometer data $G'$ and $G''$; black symbols:  reconstructed moduli $G'_f$ and $G''_f$ of the peripheral latex film. Inset: schematic cross-section of the parallel-plate geometry, showing the solvent trap and the evaporation pathway through the opening around the bottom-plate shaft.}
    \label{fig:Fig1} 
\end{figure}

The suspension is introduced in a parallel-plate geometry (diameter $40~\rm mm$, gap $500~\mu\rm m$) connected to a strain-controlled rheometer (ARES-G2, TA Instruments). 
A solvent trap is used to limit water evaporation. However, the strain-controlled configuration leaves a narrow opening around the rotating shaft of the bottom plate, through which water can slowly evaporate (see inset in Fig.~\ref{fig:Fig1}).
A strong preshear ($\dot \gamma =100~\rm s^{-1}$ for $180~\rm s$) is applied to erase the loading history, after which the linear viscoelastic moduli $G'$ and $G''$ are monitored at fixed oscillation amplitude ($\gamma_0=1\%$), either at fixed frequency ($\omega=2\pi~\rm rad.s^{-1}$) or by TRMS \cite{Mours:1994}. In the latter case, the apparent viscoelastic spectrum, $G'$ and $G''$ vs.~$\omega$, is measured over 9 to 12 discrete frequencies ranging from 0.628 to 113~rad.s$^{-1}$, approximately every $\Delta t_{\rm exp} =60~\rm s$, thanks to a superposition of small-amplitude oscillatory strains. This allows us to track the crossover of $G'$ and $G''$, while ensuring that the suspension properties do not evolve significantly over the measurement duration $\Delta t_{\rm exp}$ \cite{Mours:1994,Winter:1988}.

\begin{figure}[t!]
    \centering
    \includegraphics[width=1\linewidth]{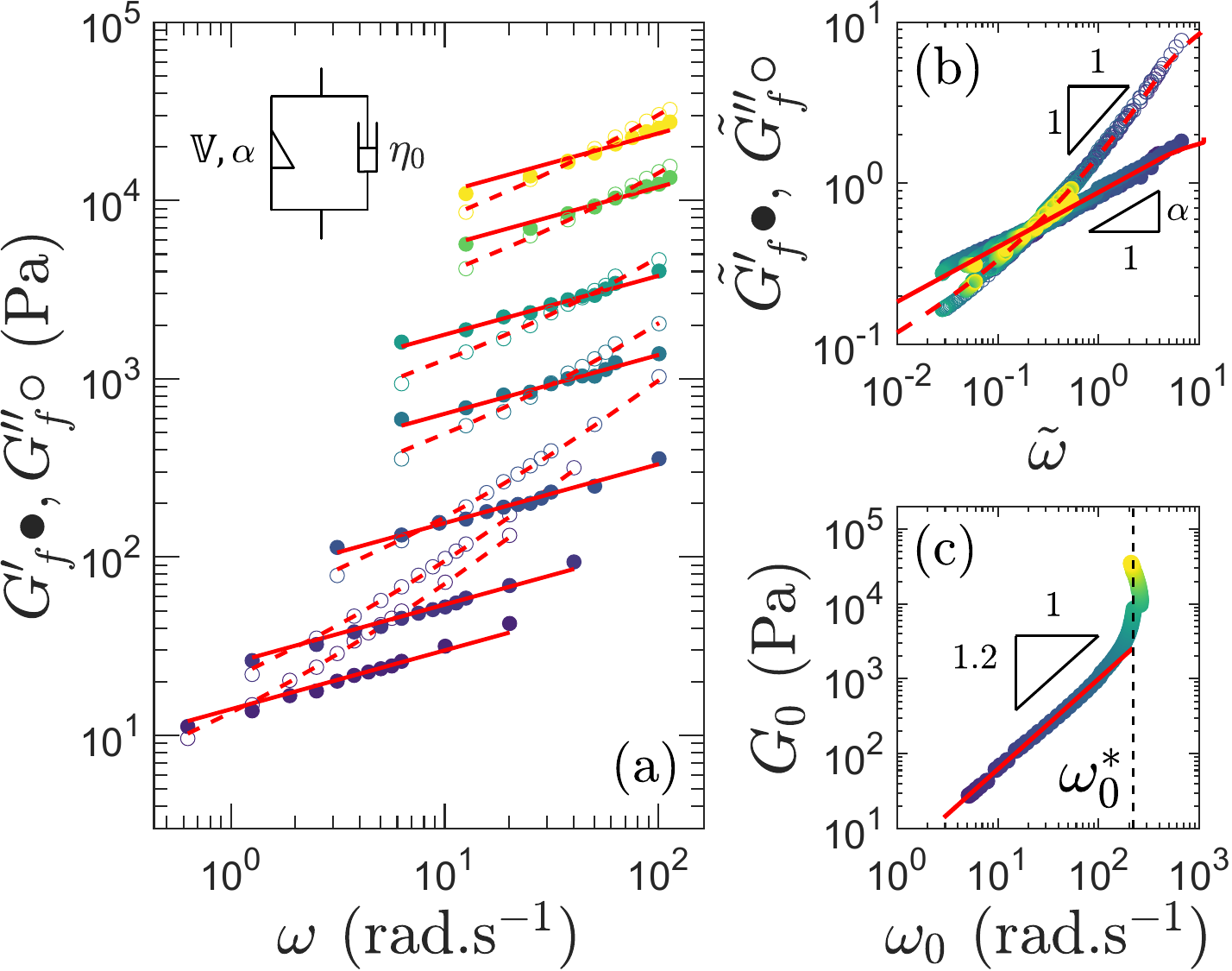}
    \caption{\label{fig:Fig2} TRMS during drying of the dense latex suspension. (a)~Frequency dependence of the film viscoelastic moduli $G'_f$ and $G''_f$ at $t = 874$, $1228$, $1840$, $2698$, $3526$, $4779$, and $6085~\rm s$ (dark blue to yellow). Red curves: two-parameter fits by the FKV model with $\alpha=0.33$ sketched in the inset [see Eq.~\eqref{eq:1}]. (b) Master curve obtained using: $\tilde G'_f=G'_f/G_0$, $\tilde G''_f=G''_f/G_0$, and $\tilde\omega = \omega/\omega_0$, with $G_0=(\mathbb{V}^{1/\alpha}/\eta_0)^{\alpha/(1-\alpha)}$ and $\omega_0=(\mathbb{V}/\eta_0)^{1/(1-\alpha)}$. Red curves: normalized FKV model $\tilde G^*= i \tilde \omega + (i \tilde \omega)^\alpha$, with $\alpha=0.33$. (c) $G_0$ vs.~$\omega_0$. Red line: power-law fit for $\omega_0 \leq 100~\rm rad.s^{-1}$, with exponent $1.20 \pm 0.02$; vertical dashed line: $\omega_0=\omega_0^*=218~\rm rad.s^{-1}$ reached at $t_c =3716~\rm s$.}
\end{figure}

\textit{Results.-} The linear viscoelastic properties, reported in Fig.~\ref{fig:Fig1}, display an apparent ``\textit{delayed stiffening}" following preshear. The suspension initially behaves as a liquid ($G' \lesssim G''$), and both moduli remain nearly constant for about $800~\rm s$, before increasing sharply as a power law with an exponent of about $4.0$. This behavior, consistently observed for sufficiently large volume fractions $\phi \gtrsim 40\%~\rm vol$ (see Fig.~S3 in the SM), differs markedly from the slow logarithmic aging typically observed in soft glasses \cite{Divoux:2011,Erwin:2011,Joshi:2018}. Visual observations reveal that this delayed increase in viscoelasticity is accompanied by the formation of a coalesced elastic film at the periphery of the geometry, whose radial thickness progressively increases while the center of the sample remains liquid (see Fig.~S4 in the SM). Particle-size measurements performed on a sample collected from the center of the geometry show an unchanged size distribution (see Fig.~S2 in the SM), indicating that particle coalescence is confined to the periphery.
This heterogeneous film formation results from slow water evaporation through the opening around the bottom-plate shaft. Crucially, when evaporation is suppressed by surrounding the sample with oil, both $G'$ and $G''$ remain constant (see Fig.~S5 in the SM), confirming that the increase in viscoelasticity is driven by drying.
Hereafter, $t$ denotes the time elapsed since cessation of preshear and thus the drying time. 

\begin{figure*}[th!]
    \centering
    \includegraphics[width=0.9\linewidth]{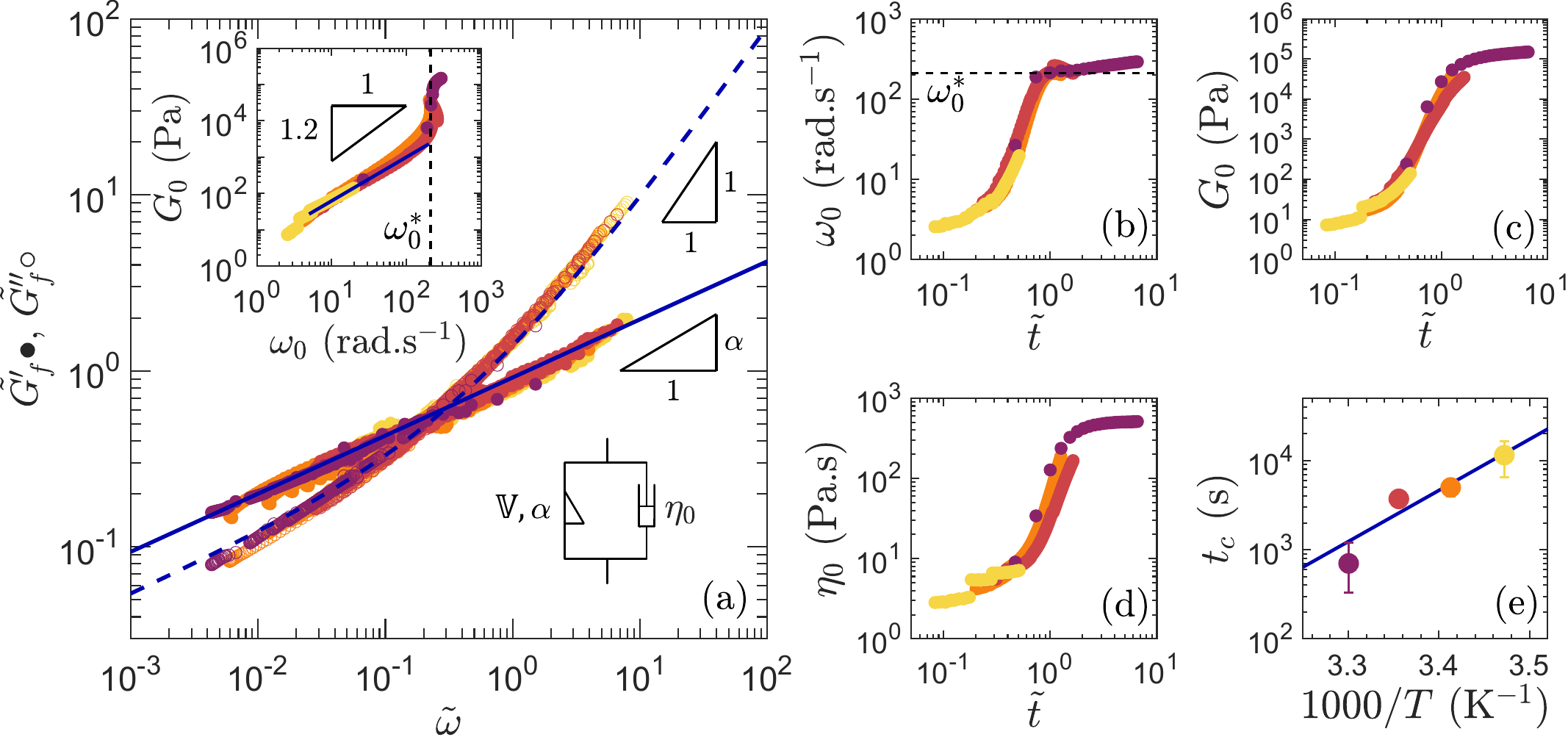}
    \caption{\label{fig:Fig3} (a) Master curve for the film viscoelastic spectra obtained at four different temperatures ($T=15,\, 20,\, 25,\, \rm and\, 30^\circ$C from yellow to purple). The blue curves correspond to the normalized FKV model with $\alpha=0.33$. Inset: $G_0$ vs.~$\omega_0$. The blue line is the best power-law fit of the data with an exponent $1.2$. The vertical dashed line highlights $\omega_0^*= 218 ~\rm rad.s^{-1}$. The sketch shows the FKV model. (b)--(d) Evolution of $\omega_0$, $G_0$, and $\eta_0$ vs.~normalized time $\tilde t=t/t_c$, where $t_c$ depends on the temperature and denotes the time that best collapses the evolutions of $\omega_0(\tilde t)$ and $G_0(\tilde t)$. (e) $t_c$ vs.~$1000/T$. The blue line corresponds to an Arrhenius behavior $t_c \propto \exp(U/RT)$, yielding $U = 110\pm 15~\rm kJ/mol$.}
\end{figure*}

The heterogeneous sample geometry implies that the measured torque contains contributions from both the liquid-like central region and the peripheral film. We therefore decompose the total complex torque $\Gamma_{\rm tot}^*$ into the contribution of the initial liquid latex suspension $\Gamma_{\rm ini}^*$, and that of the peripheral film: $\Gamma_f^* = \Gamma_{\rm tot}^* - \Gamma_{\rm ini}^*$. The corresponding complex modulus $G_{f}^*$ of the forming film is then obtained by accounting for its time-dependent radial thickness $\xi(t)$. We assume $\xi(t) \propto t^{1/2}$, consistent with the late-stage growth of consolidated layers predicted for confined directional drying of colloidal dispersions when solvent transport becomes limiting \cite{Pingulkar:2023,Dufresne:2003,Tanaka:2023}. The detailed derivation of the recalculated moduli thanks to the assumed growth law is provided in the SM. The resulting film contribution, $G'_f$ and $G''_f$, is shown in Fig.~\ref{fig:Fig1}.
 
To better resolve the evolution of the viscoelastic response, we monitor the film formation using TRMS. A selection of spectra, measured at $7$ drying times, is shown in Fig.~\ref{fig:Fig2}(a). At $t=874~\rm s$, the spectrum already exhibits $G'_f> G''_f$ at low frequencies, indicating a solid-like response of the latex suspension. 
Subsequent spectra show a similar frequency dependence, while the $G'_f$--$G''_f$ crossover shifts towards higher frequencies, and the viscoelastic modulus increases by four orders of magnitude. All spectra are accurately described by a Fractional Kelvin--Voigt (FKV) model, sketched in Fig.~\ref{fig:Fig2}(a), consisting of a dashpot of viscosity $\eta_0$ in parallel with a spring-pot. The latter is an intermediate mechanical element between a spring and a dashpot, defined by a fractional constitutive equation that links the stress $\sigma$ and the strain $\gamma$, $\sigma=\mathbb{V}{\rm d}^\alpha \gamma/{\rm d} t^\alpha$, where $\alpha$ is a dimensionless exponent ($0 \leq \alpha \leq 1$) and $\mathbb{V}$ is a ``quasi-property'' with dimension $\rm Pa.s^\alpha$ \cite{Friedrich:1999,Jaishankar:2013,Bonfanti:2020}.  
The corresponding complex modulus reads
\begin{equation} \label{eq:1}
G^*=i\eta_0\omega + \mathbb{V}(i\omega)^\alpha \, ,   
\end{equation}
An excellent description of all spectra is obtained with a  fixed $\alpha=0.33\pm0.03$, leaving $\eta_0$ and $\mathbb{V}$ as the only fitting parameters; both increase as a function of time (see Fig.~S6 in the SM). 
Moreover, the two fitting parameters define an elasticity scale $G_0=(\mathbb{V}^{1/\alpha}/\eta_0)^{\alpha/(1-\alpha)}$ and a frequency scale $\omega_0=(\mathbb{V}/\eta_0)^{1/(1-\alpha)}$. Rescaling the film viscoelastic spectra by these two quantities produces a remarkable master curve spanning three decades in dimensionless frequency and two decades in dimensionless elasticity [see Fig.~\ref{fig:Fig2}(b)]. The corresponding master curve of $\tan \delta =G''_f/G'_f$ is shown in Fig.~S7 in the SM. This master curve indicates that evolution of the viscoelastic response during film formation obeys a time-aging superposition principle \cite{Joshi:2025}, reminiscent of the time-connectivity superposition principle reported across the sol-gel transition of dilute colloidal suspensions \cite{Keshavarz:2021,MorletDecarnin:2023,Bantawa:2023}. 
The evolution of the shift factors further reveals two distinct regimes. At early times, $G_0\sim \omega_0^{1.2}$, whereas at later times $G_0$ continues to increase while $\omega_0$ approaches a plateau, $\omega_0 \simeq \omega_0^*= 218~\rm rad.s^{-1}$  [see Fig.~\ref{fig:Fig2}(c)]. This crossover signals a radical change in the dynamics of film formation, as discussed below.  

\begin{figure*}[t!]
    \centering
    \includegraphics[width=0.85\linewidth]{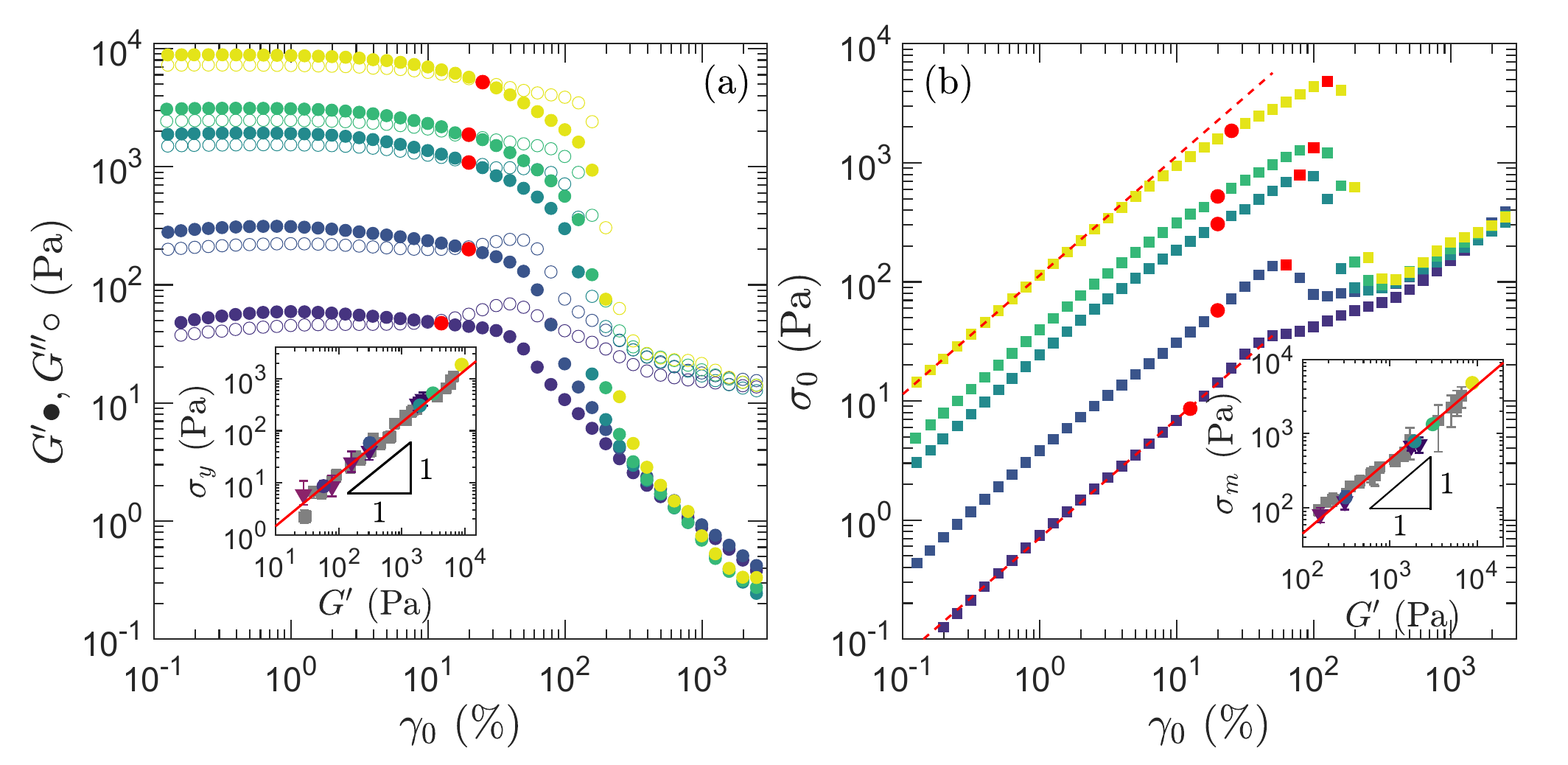}
    \caption{\label{fig:Fig4} Non-linear rheology of the dense latex suspension under large amplitude oscillatory shear. The moduli and stresses are apparent quantities calculated from the total measured torque, without correcting for the heterogeneous sample geometry. (a) $G'$ and $G''$ vs.~strain amplitude $\gamma_0$ measured at $\omega=2\pi~\rm rad.s^{-1}$ at five drying times: $t=1088, 1751, 3037, 4200, \rm{and}\, 5903\, \rm s$ (dark blue to yellow, same color code as in Fig.~\ref{fig:Fig2}). Red dots indicate the yield points at $\gamma_y$, defined by the crossover of $G'$ and $G''$. Inset: Yield stress, i.e., the stress amplitude $\sigma_y$ at $\gamma_y$, vs.~$G'$; the red line is the best linear fit, $\sigma_y=0.15 G'$. Blue-to-purple triangles correspond to tests conducted at fixed drying time ($t=446~\rm s$) and different temperatures (see Fig.~S9 in the SM), while gray squares correspond to samples subjected to repeated cycles of drying and yielding at $T=25^\circ$C. (b) Stress amplitude $\sigma_0$ vs.~$\gamma_0$ for the same strain sweeps as in (a). Red dashed lines indicate a linear scaling; red dots indicate the yield points and red squares mark the stress maxima $\sigma_m$, whose dependence on $G'$ is shown as an inset. The red line is the best linear fit, $\sigma_m= 0.45 G'$.}
\end{figure*}

The robustness of this time-aging superposition is demonstrated at three other temperatures, $T=15$, $20$, and $30^\circ \rm C$ [see Fig.~\ref{fig:Fig3}(a)]. The corresponding master curves nearly superimpose (see also Fig.~S8 in the SM), indicating that the spectral shape is independent of temperature over the range investigated. In contrast, temperature strongly affects the evolution of the shift factors $G_0$ and $\omega_0$, as well as $\eta_0$. These quantities collapse onto master curves when plotted against the normalized time $\tilde t=t/t_c$, where $t_c$ is the characteristic time yielding the best overlap of the data [see Fig.~\ref{fig:Fig3}(b)-(d)]. The characteristic time $t_c$ shows an Arrhenius-like temperature dependence [see Fig.~\ref{fig:Fig3}(e)], with an apparent activation energy $U = 110~\pm 15~\rm kJ/mol$, compatible with the energy scale associated with strong hydrophobic interactions \cite{Reynolds:1974,Mihajlovic:2017}. This result suggests that the particles progressively overcome their steric repulsion, allowing attractive hydrophobic interactions to drive film formation. 

Finally, we probe the non-linear response of the latex suspension at different stages of the film formation to characterize the evolution of its yielding behavior. 
Unlike the linear viscoelastic measurements, the nonlinear data are reported without correcting for the peripheral film geometry, since both the liquid core and the film contribute to the measured response in a strain-dependent manner.  
Figure~\ref{fig:Fig4}(a) shows strain sweeps conducted at a fixed frequency ($\omega=2\pi~\rm rad.s^{-1})$ for several sample ages. At low strain amplitude, the latex suspension displays a solid-like response ($G'>G''$), followed by yielding at $\gamma_0=\gamma_y$, defined by the crossover of $G'$ and $G''$, and a liquid-like response ($G'<G''$) at strain amplitudes above $\gamma_y$. 
The yield strain remains approximately constant, $\gamma_y \simeq 20\%$, over the range of drying times investigated, whereas the yield stress $\sigma_y$, defined as the stress amplitude at $\gamma_y$, increases linearly with the elastic modulus $G'$, with a prefactor of about 0.15, independently of the drying time, temperature, and shear history [see inset in Fig.~\ref{fig:Fig4}(a)].

The stress amplitude $\sigma_0(\gamma_0)$ provides additional insights regarding the microstructural changes occurring during the film formation [Fig.~\ref{fig:Fig4}(b)]. At early times after preshear (here $t=1088$~s~$ < t_c$), the stress amplitude $\sigma_0$ initially increases linearly with strain and then deviates from this scaling above $\gamma_0 \simeq 40\%$. With increasing drying time, the deviation occurs at lower strain values, around $10\%$, and is accompanied by an increasingly pronounced stress overshoot. 
This transition from a purely strain-softening response to a combination of strain-softening and stress overshoot is indicative of a ductile-to-brittle transition and reminiscent of results obtained in soft glasses upon increasing attractive forces \cite{Pham:2006,Pham:2008}. 
Moreover, the stress maximum $\sigma_m$ characterizing the stress overshoot scales linearly with the sample elastic modulus $G'$ [see inset in Fig.~\ref{fig:Fig4}(b)], corresponding to a characteristic strain of order unity, about ten times larger than the yield strain $\gamma_y$.

\textit{Discussion and conclusion.-} The \textit{delayed} increase in viscoelasticity results from drying-induced film formation rather than the usual aging dynamics observed during recovery after preshear. At the microscale, we propose that the densely packed latex particles interact predominantly through their repulsive coronas. As water evaporates, the particles are brought closer together, allowing their polymer shells to interpenetrate \cite{LaraPena:2021,Saisavadas:2023} and hydrophobic interactions to develop, as supported by the estimation of the activation energy in Fig.~\ref{fig:Fig3}. These interactions promote particle coalescence and the formation of a continuous peripheral film, accounting for the pronounced increase in viscoelastic moduli and the transition from ductile to brittle yielding. 

The time-aging superposition provides further insight into the dynamics of film formation. The characteristic time $t_c$, beyond which the frequency shift factor $\omega_0$ reaches a plateau, suggests a crossover toward arrested particle dynamics. A similar behavior was reported in gels of soft hairy colloids, where the plateau of $\omega_0$ marks the onset of contact-driven aging through particle sintering \cite{Bauland:2024}. By analogy, we propose that beyond $t_c$, particle rearrangements become strongly hindered and the mechanical evolution is increasingly governed by the strengthening and coalescence of interparticle contacts. Unlike these colloidal gels, however, which exhibit $\sigma_y\propto\sqrt{G'}$ as contacts age, the present latex system displays $\sigma_y\propto G'$, indicative of a distinct mechanical evolution associated with the progressive coalescence of particles into a continuous film.
Whether this crossover should be viewed as a gel-like transition or as a rigidity transition in a dense assembly of increasingly attractive particles remains unclear. The nearly frequency-independent loss factor $\tan\delta$ at $\tilde\omega\ll 1$ (see Fig.~S8 in the SM) is reminiscent of the critical-like mechanical response observed at the gel point \cite{Winter:1999}, while rigidity transitions have also been predicted for dense assemblies of attractive particles \cite{Lois:2008,Koeze:2018,Zhang:2019}. Interestingly, the FKV exponent $\alpha$ decreases slightly with increasing particle volume fraction \cite{Exponent}, suggesting that $\alpha$ may also reflect features of the underlying microstructure, as previously proposed for particulate gels \cite{Larsen:2008,Keshavarz:2021,Bantawa:2023,Bauland:2024}. Distinguishing between these scenarios and elucidating the microscopic origin of $\alpha$ will require direct structural characterization during drying.


\begin{acknowledgments}
The authors are particularly grateful to J.-B.~Salmon for insightful discussions on drying dynamics. They also acknowledge S.~Dankic-Cottrino for providing access to the master sizer, G.~Legrand for his critical review of the manuscript, as well as G.P.~Baeza, J.~Bauland, and J.~Blin for fruitful discussions. The authors gratefully acknowledge Saint-Gobain for financial support. This work was supported by the LABEX iMUST of the University of Lyon (ANR-10-LABX-0064), created within the ``Plan France 2030" set up by the French government and managed by the French National Research Agency (ANR), and by the ANRT (Association Nationale de la Recherche et de la Technologie) through a CIFRE fellowship granted to J.~Henry. 
\end{acknowledgments}


\providecommand{\noopsort}[1]{}\providecommand{\singleletter}[1]{#1}%

\clearpage

\begin{widetext}

\begin{center}
{\Large --Supplemental Materials--}
\end{center}

\section*{Scanning electron microscopy of the redispersible polymer powder}

Figure~\ref{fig:FigS1} shows representative scanning electron microscopy (SEM) images of the commercial redispersible polymer powder (Saint-Gobain) prior to dispersion in water. Images were acquired using a JEOL IT800HL microscope at an accelerating voltage of $3.00~\rm kV$ and a working distance of about $2~\rm mm$. The powder contains spherical latex particles with diameters of about a few microns together with calcium carbonate crystals and clay platelets used as stabilizing additives. 
\begin{figure}[h!]
    \centering
    \includegraphics[width=0.95\linewidth]{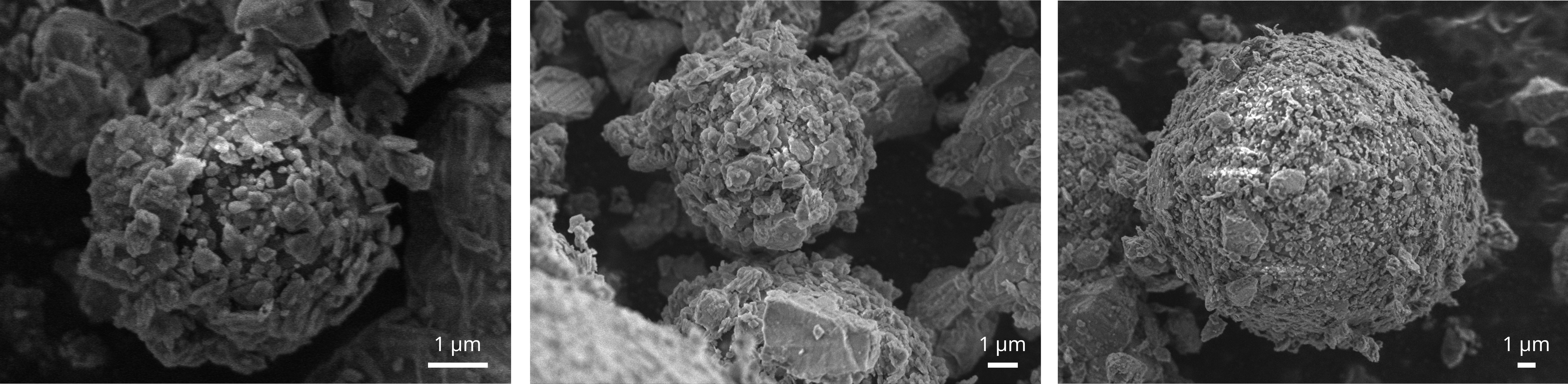}
    \caption{Representative SEM images of the redispersible polymer powder. Spherical objects correspond to latex particles, while the debris surrounding them include crystals of calcium carbonate and clay platelets.}
    \label{fig:FigS1} 
\end{figure}

\vspace{-0.9cm}

\section*{Particle-size distribution}

Figure~\ref{fig:FigS1b} compares the particle size distributions of a freshly prepared suspension and of a sample collected from the center of the parallel-plate geometry after the drying experiment reported in Fig.~1 of the main text. The distributions are nearly identical. Thus, no detectable particle coalescence occurs in the central region of the parallel plates. This indicates that a liquid-like suspension with the same composition as the initial suspension lies at the center of the parallel-plate geometry, while a coalesced film grows at its periphery.   
\begin{figure}[h!]
    \centering
    \includegraphics[width=0.4\linewidth]{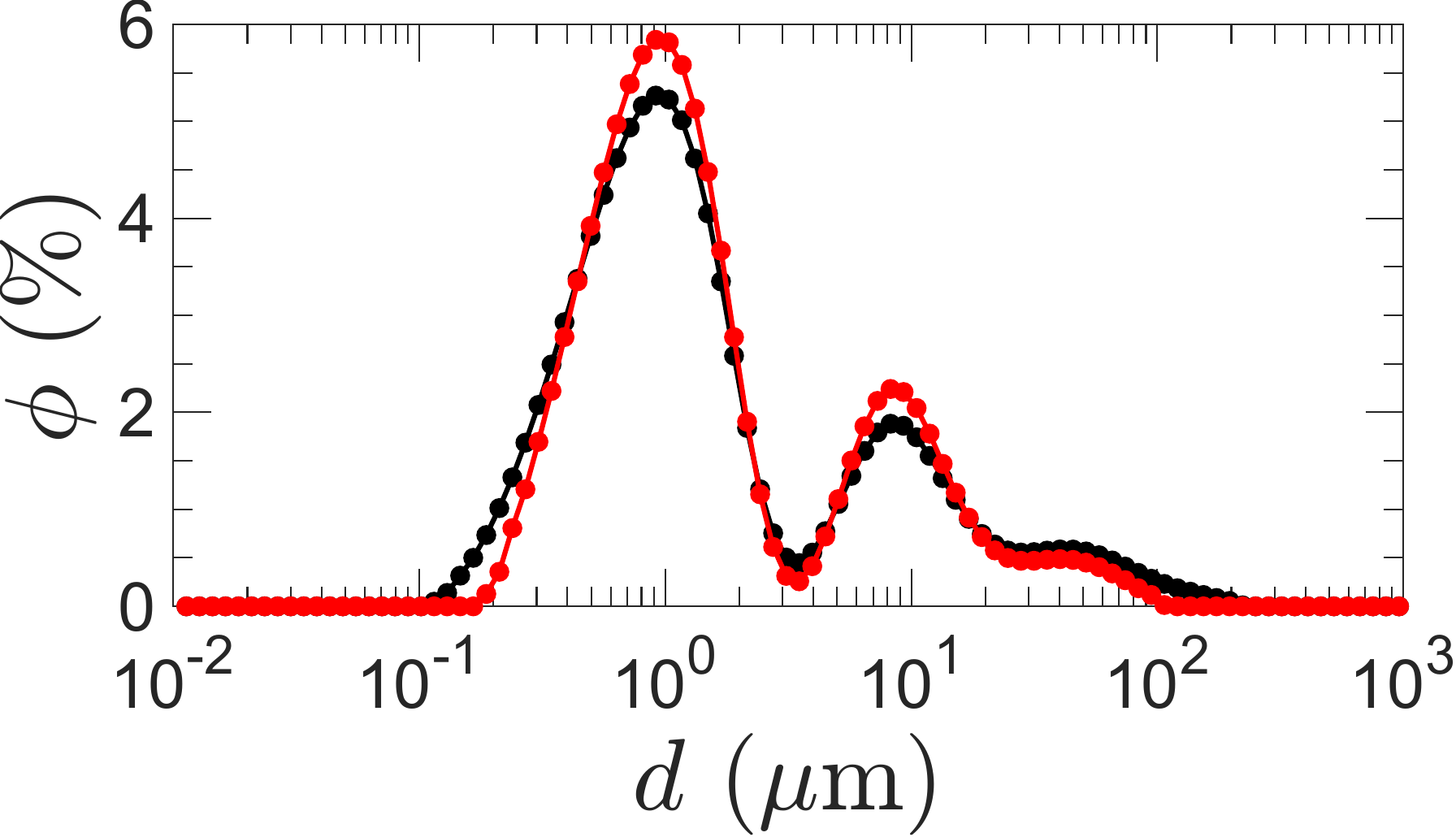}
    \caption{Particle-size distributions measured by laser diffraction (Mastersizer 2000, Malvern) before loading the suspension (black) and after the drying experiment (red). The latter sample was collected from the center of the geometry, away from the peripheral film.}
    \label{fig:FigS1b} 
\end{figure}

\clearpage

\section*{Influence of particle volume fraction on film formation}

Figure~\ref{fig:VolumeFraction} shows the evolution of the linear viscoelastic properties for latex suspensions with particle volume fractions ranging from $20.8~\%~\rm vol$ to $67.6~\%~\rm vol$. For $\phi \lesssim 40\%$, both $G'$ and $G''$ remain nearly constant over the experimental time window. At higher volume fractions, a delayed and pronounced increase in viscoelasticity is observed, with an earlier onset as $\phi$ increases. These results indicate that film formation within the experimental time window requires sufficiently concentrated suspensions, consistent with the increasing proximity of neighboring particles facilitating coalescence during drying. 

 \begin{figure}[h!]
     \centering
     \includegraphics[width=0.75\linewidth]{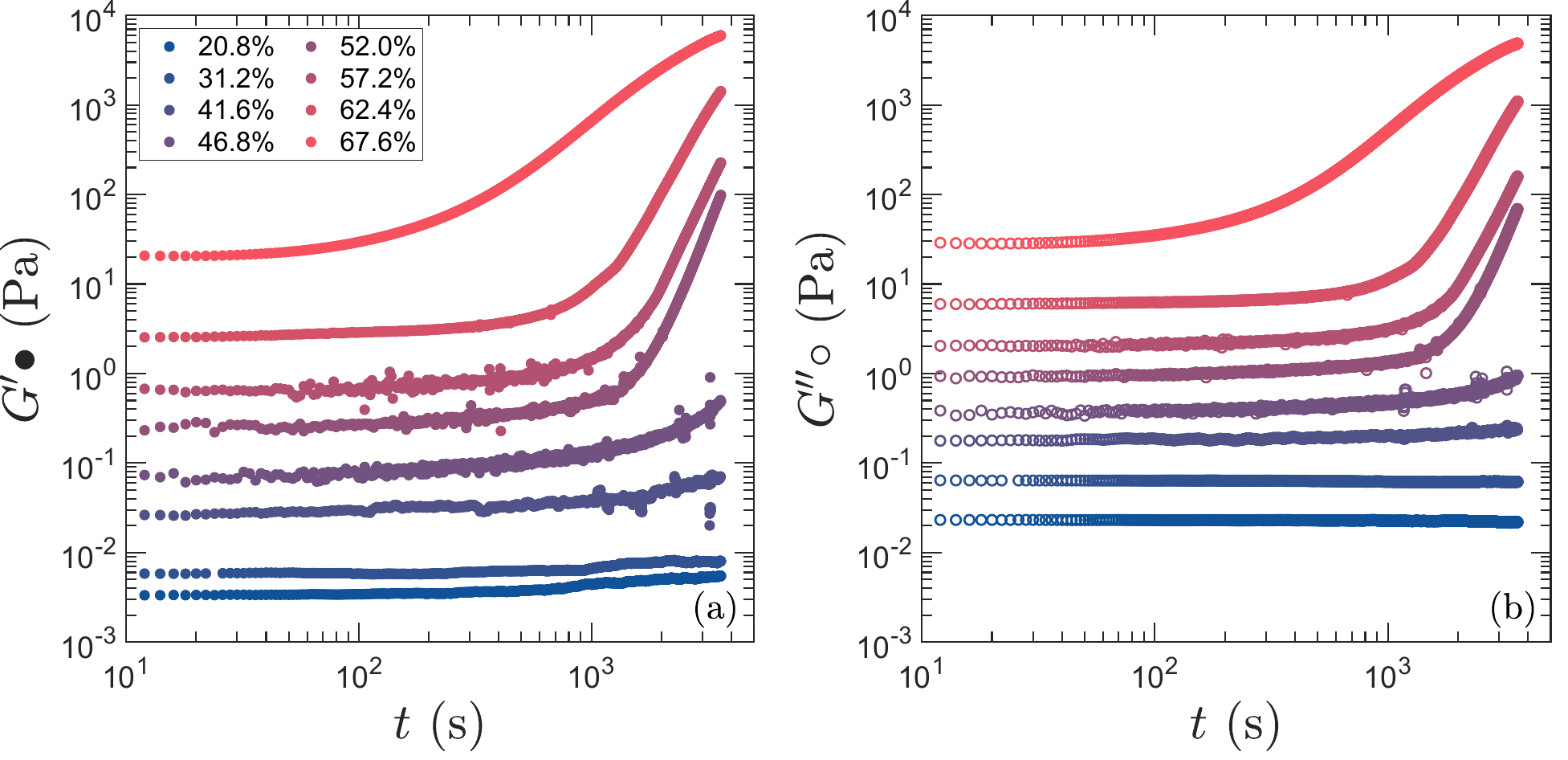}
     \caption{Evolution of (a) $G'$ and (b) $G''$ after flow cessation for latex suspensions with particle volume fractions ranging from $20.8$ to $67.6\%$. The measurements were performed at $\omega=2\pi~\rm rad.s^{-1}$ and $\gamma_0=2\%$, following a preshear at $\dot \gamma=100~\rm s^{-1}$ for $180~\rm s$. Colors indicate particle volume fraction. Experiments conducted at $T=25^\circ$C.}
     \label{fig:VolumeFraction}
 \end{figure}

\section*{Direct observation of peripheral latex film formation}

Figure~\ref{fig:SI_Rheo_Residual_Film} provides visual evidence of the spatially heterogeneous film formation observed during drying. After approximately $4500~\rm s$ at $T=25^\circ\rm C$, a coherent elastic film is present at the periphery of the parallel-plate geometry, while the central region remains liquid. Upon separating the plates, the peripheral film stretches and eventually ruptures [Fig.~\ref{fig:SI_Rheo_Residual_Film}(a)--(d)]. Removing the remaining liquid suspension reveals a continuous annular film [Fig.~\ref{fig:SI_Rheo_Residual_Film}(e),(f)], with a radial thickness of approximately $0.5$--$1~\rm mm$ depending on the duration of the experiment. These observations support the annular geometry assumed when reconstructing the viscoelastic contribution of the forming film.

 \begin{figure}[h!]
    \centering    
    \includegraphics[width=0.95\linewidth]{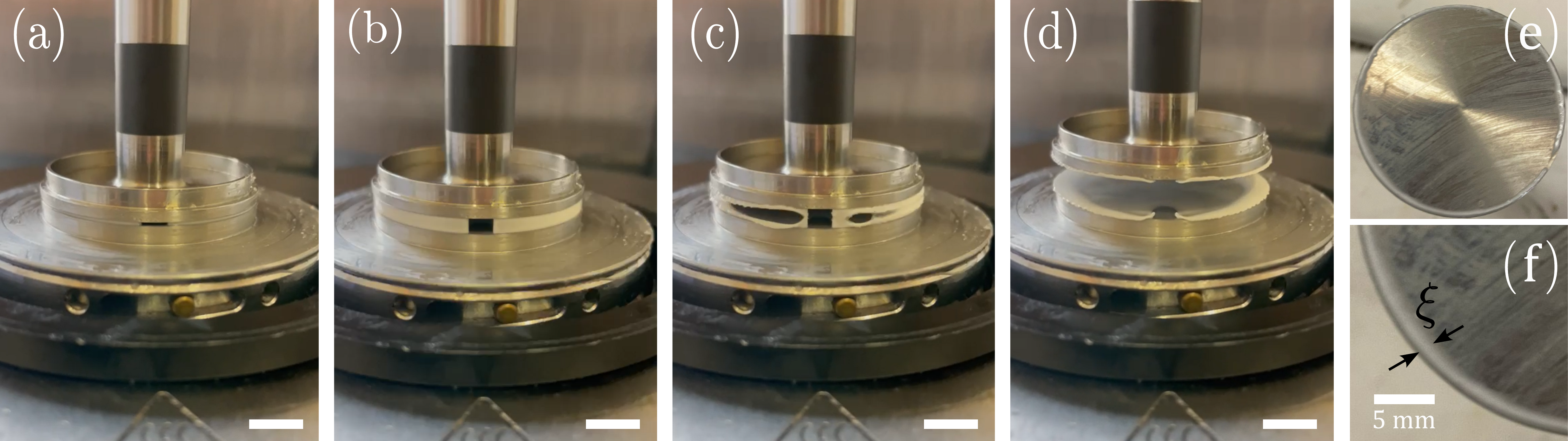}
    \caption{Photographs of the peripheral latex film formed after $4500~\rm s$ of drying at $T=25^\circ\rm C$. (a)--(d) Progressive deformation and rupture of the film as the upper plate is lifted. White scale bar: $10~\rm mm$. (e),(f) Close-up views of the annular film remaining after removal of the liquid suspension. The radial thickness $\xi$ of the film is about $1~\rm mm$.} 
\label{fig:SI_Rheo_Residual_Film}
\end{figure}

\clearpage

\section*{Control experiments: role of solvent evaporation}

To establish the role of water evaporation in the increase of viscoelasticity, we performed time sweeps on identical latex suspensions ($\phi=57.2\%$) under three experimental conditions (Fig.~\ref{fig:Rheo_Conditions_Film_Formation}). When the sample is completely isolated from ambient air by a layer of mineral oil, both $G'$ and $G''$ remain nearly constant for at least $10^4~\rm s$ [Fig.~\ref{fig:Rheo_Conditions_Film_Formation}(a)]. By contrast, the standard configuration used in the main text, which involves a water-filled solvent trap but retains a narrow opening around the rotating shaft of the bottom plate, allows slow evaporation and produces a delayed increase in both moduli [Fig.~\ref{fig:Rheo_Conditions_Film_Formation}(b)]. Removing the solvent trap promotes faster evaporation and substantially accelerates this increase [Fig.~\ref{fig:Rheo_Conditions_Film_Formation}(c)]. These controls demonstrate that the evolution of viscoelasticity is driven by water evaporation rather than spontaneous aging of the latex suspension.

 \begin{figure}[h!]
    \centering
    \includegraphics[width=0.95\linewidth]{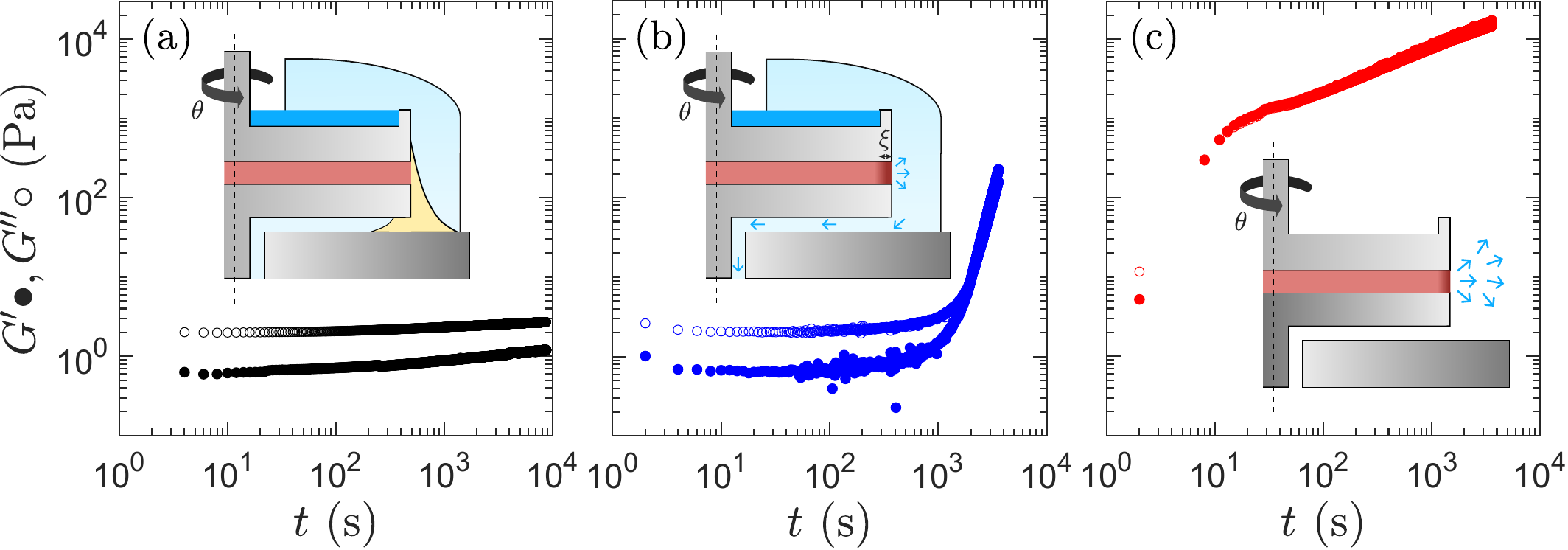}
    \caption{Evolution of $G'$ and $G''$ following preshear for identical latex suspensions ($\phi=57.2\%$) under three evaporation conditions: (a) evaporation suppressed by a layer of mineral oil surrounding the sample, (b) standard configuration with a water-filled solvent trap and a narrow opening around the bottom-plate shaft as in the main text, and (c) no solvent trap. Insets illustrate the corresponding configurations. Measurements were performed at $\omega=2\pi~\rm rad\,s^{-1}$ and $\gamma_0=1\%$, at $T=25^\circ\rm C$. The laboratory relative humidity was approximately $45\%$.}
    \label{fig:Rheo_Conditions_Film_Formation} 
\end{figure}

\vspace{-0.9cm}

\section*{Computation of the viscoelastic response of the peripheral film}

\subsection*{Torque decomposition}

 \begin{figure}[b!]
    \centering
    \includegraphics[width=0.2\linewidth]{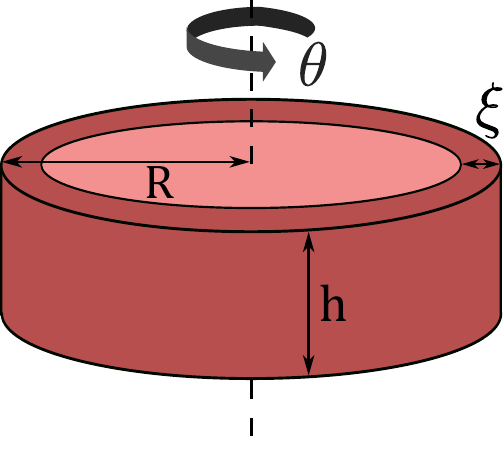}
    \label{fig:Geometry_Scheme} 
\end{figure}

During drying, the sample is modeled as a liquid-like central cylinder surrounded by a coalesced viscoelastic film of radial thickness $\xi(t)$, as illustrated in Figs.~\ref{fig:SI_Rheo_Residual_Film}(f) and \ref{fig:Rheo_Conditions_Film_Formation}(b), and sketched below without the parallel plate geometry for clarity. The film is assumed to be axisymmetric, with a uniform radial thickness at each instant. In the linear regime, the total complex torque is the sum of the contributions from the central suspension of latex particles and the peripheral latex film:
\begin{equation}
\Gamma_{\rm tot}^*
=
\frac{2\pi\theta^*}{h}
\left[
\int_0^{R-\xi}G_{\rm liq}^*r^3\,dr
+
\int_{R-\xi}^{R}G_f^*r^3\,dr
\right],
\end{equation}
where $R$ and $h$ are the plate radius and gap, respectively, $\theta^*$ is the complex angular-displacement amplitude, $G_{\rm liq}^*$ is the complex modulus of the central suspension, and $G_f^*$ is that of the peripheral latex film. 

Assuming that the central suspension retains its initial linear viscoelastic response, $G_{\rm liq}^*\simeq G_{\rm ini}^*$, integration yields
\begin{equation}
\Gamma_{\rm tot}^*
=
\frac{\pi\theta^*}{2h}
\left[
G_{\rm ini}^*(R-\xi)^4
+
G_f^*\left(R^4-(R-\xi)^4\right)
\right].
\label{eq:TorqueFilm}
\end{equation}

The reference torque of the initial suspension occupying the entire geometry is
\begin{equation}
\Gamma_{\rm ini}^*
=
\frac{\pi\theta^*R^4}{2h}G_{\rm ini}^*.
\end{equation}

Subtracting this reference torque from Eq.~\eqref{eq:TorqueFilm} gives

\begin{equation}
\Gamma_{\rm tot}^*-\Gamma_{\rm ini}^*
=
\frac{\pi\theta^*}{2h}
\left(G_f^*-G_{\rm ini}^*\right)
\left[R^4-(R-\xi)^4\right].
\end{equation}

The intrinsic complex modulus of the peripheral film therefore reads
\begin{equation}
G_f^* =
G_{\rm ini}^*
+
\frac{2h}{\pi\theta^*}
\frac{\Gamma_{\rm tot}^*-\Gamma_{\rm ini}^*}
{R^4-(R-\xi)^4}.
\label{eq:FilmCorrection}
\end{equation}

Equivalently, introducing the apparent complex modulus $G_{\rm app}^*$ calculated by the rheometer from the total torque,
\begin{equation}
G_f^*
=
G_{\rm ini}^*
+
\left(G_{\rm app}^*-G_{\rm ini}^*\right)
\frac{R^4}{R^4-(R-\xi)^4}.
\label{eq:FilmCorrectionApp}
\end{equation}
Thus, computing the film modulus requires an estimate of the time-dependent radial thickness of the film $\xi(t)$.

\subsection*{Estimate of the film growth law}

The growth of a dense colloidal layer at an evaporation front depends on particle transport and solvent permeability. Two regimes are commonly reported. For dilute suspensions in which solvent transport remains relatively unimpeded, the layer thickness can increase approximately linearly with time \cite{wallenstein2011theory,lidon2014dynamics,abe2024dynamics,goehring2010solidification,sui2019transport}. In contrast, a sublinear growth law, typically $\xi\propto t^{1/2}$, has been observed when solvent transport becomes limited by the formation of a dense colloidal or polymer-rich layer \cite{Pingulkar:2023,Tanaka:2023}.

The present suspension is initially highly concentrated, and the formation of a coalesced peripheral film is expected to reduce water permeability near the evaporation front. This is consistent with the skinning regime described in the film-formation framework of Routh and Russel \cite{routh1999process,routh2001deformation}. We therefore assume that the radial thickness of the peripheral film grows according to
\begin{equation}
\xi(t)=\xi_{\rm ref}\left(\frac{t}{t_{\rm ref}}\right)^{1/2},
\label{eq:deltaGrowth}
\end{equation}
where $t=0$ corresponds to the end of the preshear step, and $\xi_{\rm ref}$ denotes the film width at a reference drying time $t_{\rm ref}$. This expression assumes that evaporation begins immediately after flow cessation. Based on Fig.~\ref{fig:SI_Rheo_Residual_Film}, we estimate that the peripheral film reaches a radial thickness of $\xi_{\rm ref}\simeq 1~\rm mm$ at $t_{\rm ref}=4500~\rm s$.

\clearpage

\section*{Time evolution of the FVK model parameters and rescaling factors}

Figure~\ref{fig:FigS2} shows the time dependence of the parameters obtained by fitting the linear viscoelastic spectra at $T=25^\circ\rm C$ with the Fractional Kelvin-Voigt (FKV) model [see Eq.~(1) in the main text]. Since the exponent $\alpha$ is fixed to a constant value, $\alpha=0.33$, the only two fitting parameters of the FKV model are the viscosity $\eta_0$ [Fig.~\ref{fig:FigS2}(a)] and the quasi-property $\mathbb{V}$ [Fig.~\ref{fig:FigS2}(b)]. These parameters are used to compute a frequency scale $\omega_0=(\mathbb{V}/\eta_0)^{1/(1-\alpha)}$ [Fig.~\ref{fig:FigS2}(c)] and an elasticity scale $G_0=(\mathbb{V}^{1/\alpha}/\eta_0)^{\alpha/(1-\alpha)}$ [Fig.~\ref{fig:FigS2}(d)]. 
These four parameters $\omega_0$, $G_0$, $\eta_0$, and $\mathbb{V}$ change only weakly at early drying times ($t \lesssim 1500~\rm s$) before increasing rapidly during film formation. In particular, $\omega_0$ eventually approaches a plateau $\omega_0^*\simeq218~\rm rad.s^{-1}$, at the characteristic time $t_c\simeq3716~\rm s$, while $G_0$ continues to increase.

\begin{figure}[h!]
    \centering
    \includegraphics[width=0.50\linewidth]{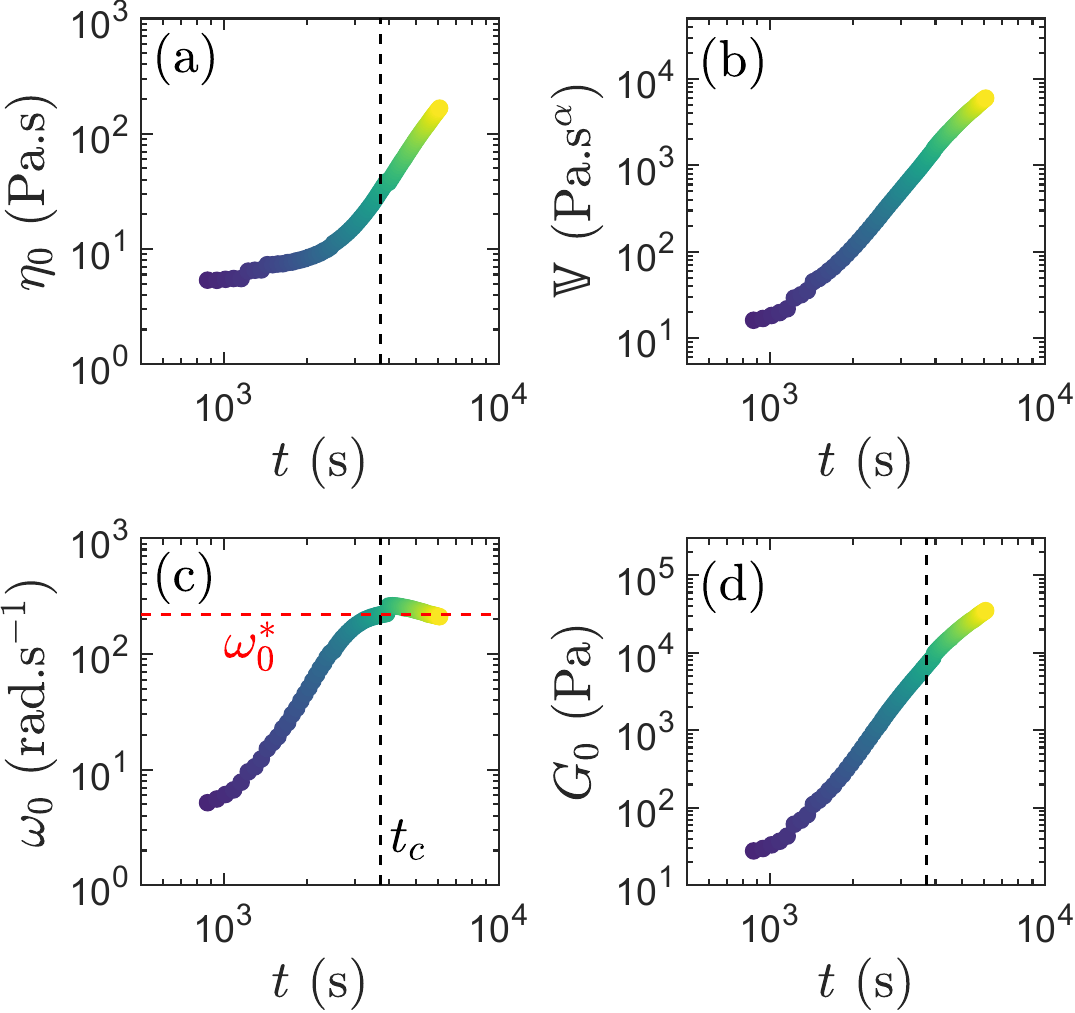}
    \caption{Time evolution of (a) $\eta_0$, (b) $\mathbb{V}$, (c) $\omega_0$, and (d) $G_0$ during film formation at $T=25^\circ\rm C$. The first two parameters are obtained from FKV fits, whereas the latter two are computed from $\eta_0$, $\mathbb{V}$, and $\alpha=0.33$. Colors indicate the drying time using the same color scale as in Figs.~2 and 4 in the main text. The vertical dashed line corresponds to $t=t_c \simeq3716~\rm s$, beyond which $\omega_0$ is approximately constant, i.e., $\omega_0\simeq \omega_0^*=218~\rm rad.s^{-1}$, as highlighted in (c) by the red dashed horizontal line. }
    \label{fig:FigS2} 
\end{figure}

\clearpage

\section*{Frequency dependence and superposition of the loss factor}

Figure~\ref{fig:FigS3}(a) shows the loss factor $\tan\delta=G_{f}''/G_{f}'$ corresponding to the film viscoelastic spectra reported in Fig.~2(a) in the main text. As film formation proceeds, the loss factor decreases and becomes less frequency-dependent. Upon rescaling the frequency by $\omega_0$, all curves collapse onto a master curve shown in Fig.~\ref{fig:FigS3}(b), consistent with the time-aging superposition of the viscoelastic moduli.

\begin{figure*}[h!]
    \centering
    \includegraphics[width=0.6\linewidth]{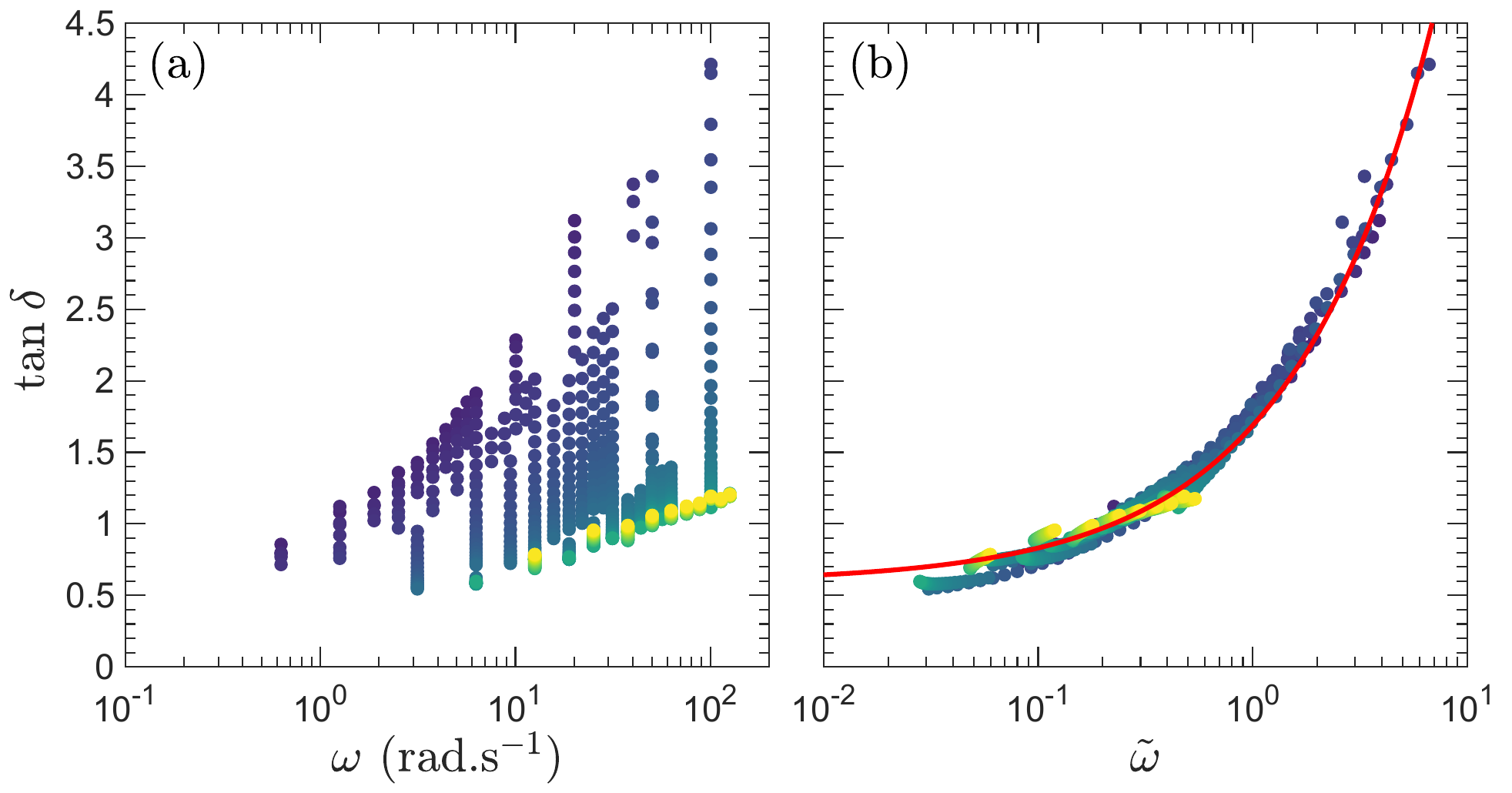}
    \caption{(a)~Loss factor $\tan \delta = G_f''/G_f'$ vs.~$\omega$ at different drying times. Same experiment as in Fig.~2(a) in the main text. (b)~Same data as in (a) plotted against normalized frequency $\tilde \omega=\omega/\omega_0$. Colors code for the time elapsed since the stop of the preshear and match the colors used in Fig.~2(a) in the main text. The red curve represents the normalized FKV model with $\alpha=0.33$ [see Eq.~\eqref{eq:tandeltaFKV}.}
    \label{fig:FigS3} 
\end{figure*}

\vspace{-1cm}
\section*{Master curve of the loss factor for different temperatures}

Figure~\ref{fig:FigS4} extends the loss-factor superposition to measurements performed at different temperatures. The collapse of $\tan\delta$ as a function of normalized frequency demonstrates that the shape of the viscoelastic spectra is nearly independent of temperature over the investigated range. Moreover, the FKV model, which predicts
\begin{equation}
\tan\delta=\tan\left(\alpha\frac{\pi}{2}\right)+\frac{\tilde{\omega}^{1-\alpha}}{\cos\left(\alpha\frac{\pi}{2}\right)},
\label{eq:tandeltaFKV}
\end{equation}
provides a very good fit of the data over the whole range of rescaled frequency (see red lines in Figs.~\ref{fig:FigS3} and \ref{fig:FigS4}).

\begin{figure*}[h!]
    \centering
    \includegraphics[width=0.35\linewidth]{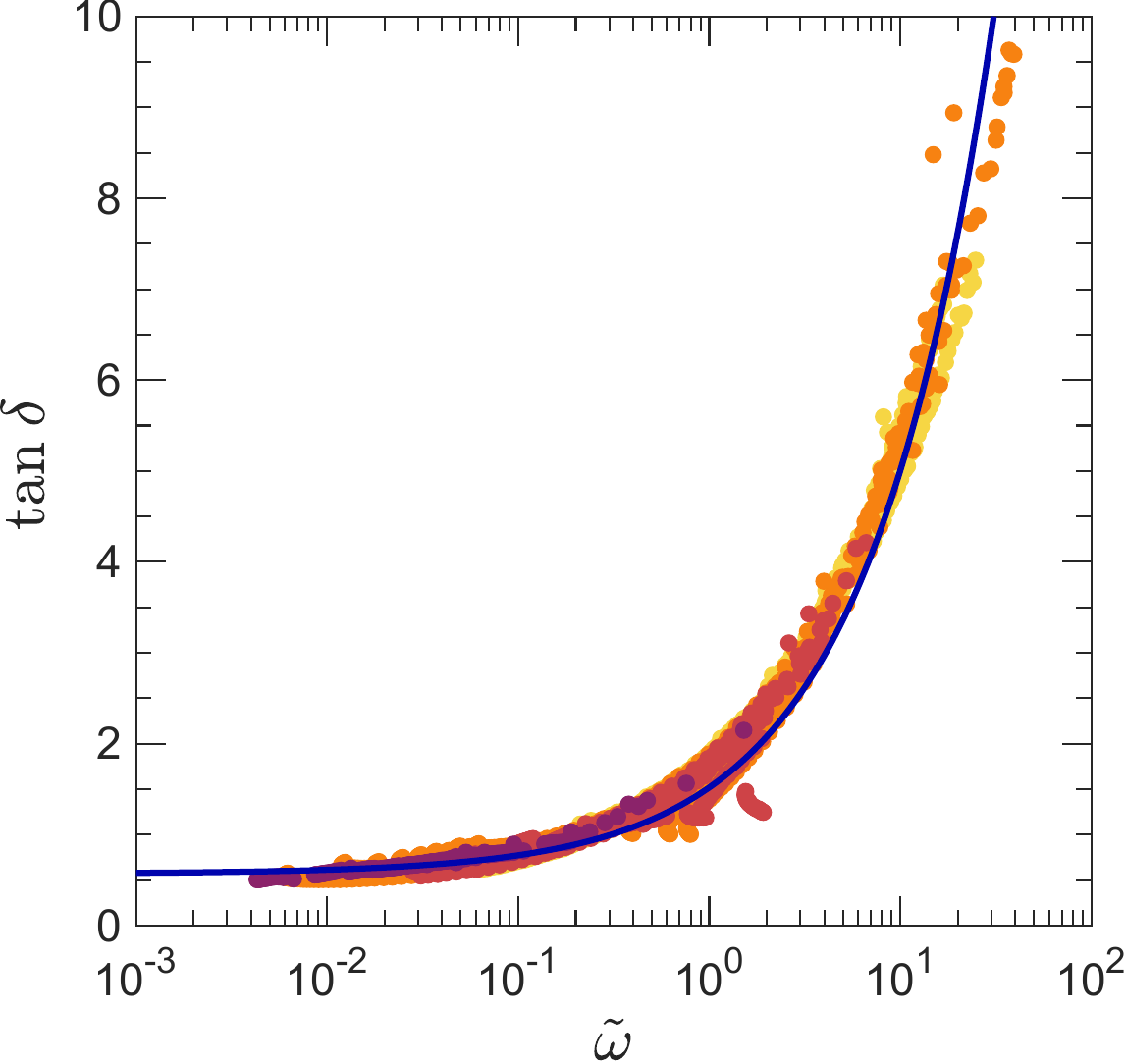}
    \caption{Loss factor $\tan \delta$ vs.~normalized angular frequency $\tilde \omega=\omega/\omega_0$ at different temperatures. Data correspond to the viscoelastic spectra reported in Fig.~3(a) of the main text. Colors indicate temperature and match the colors used in Fig.~3 in the main text. The green curve represents the normalized FKV model with $\alpha=0.33$ [see Eq.~(\ref{eq:tandeltaFKV}).}
    \label{fig:FigS4} 
\end{figure*}

\clearpage

 \section*{Nonlinear rheology at different temperatures}

 Figure~\ref{fig:FigS5} shows strain sweeps conducted on freshly prepared latex suspensions at a fixed drying time ($t=446~\rm s$) and six temperatures: $T=24, 25, 26, 28, 29, \rm and\, 30^\circ$C. The apparent yield stress $\sigma_y$, defined by the crossover of $G'$ and $G''$, and the maximum stress amplitude $\sigma_m$ are included in the corresponding scaling plots in Fig.~4 of the main text. As in Fig.~4, the moduli and stresses are calculated from the total measured torque without correcting for the heterogeneous sample geometry.

 \begin{figure*}[h!]
     \centering
     \includegraphics[width=0.7\linewidth]{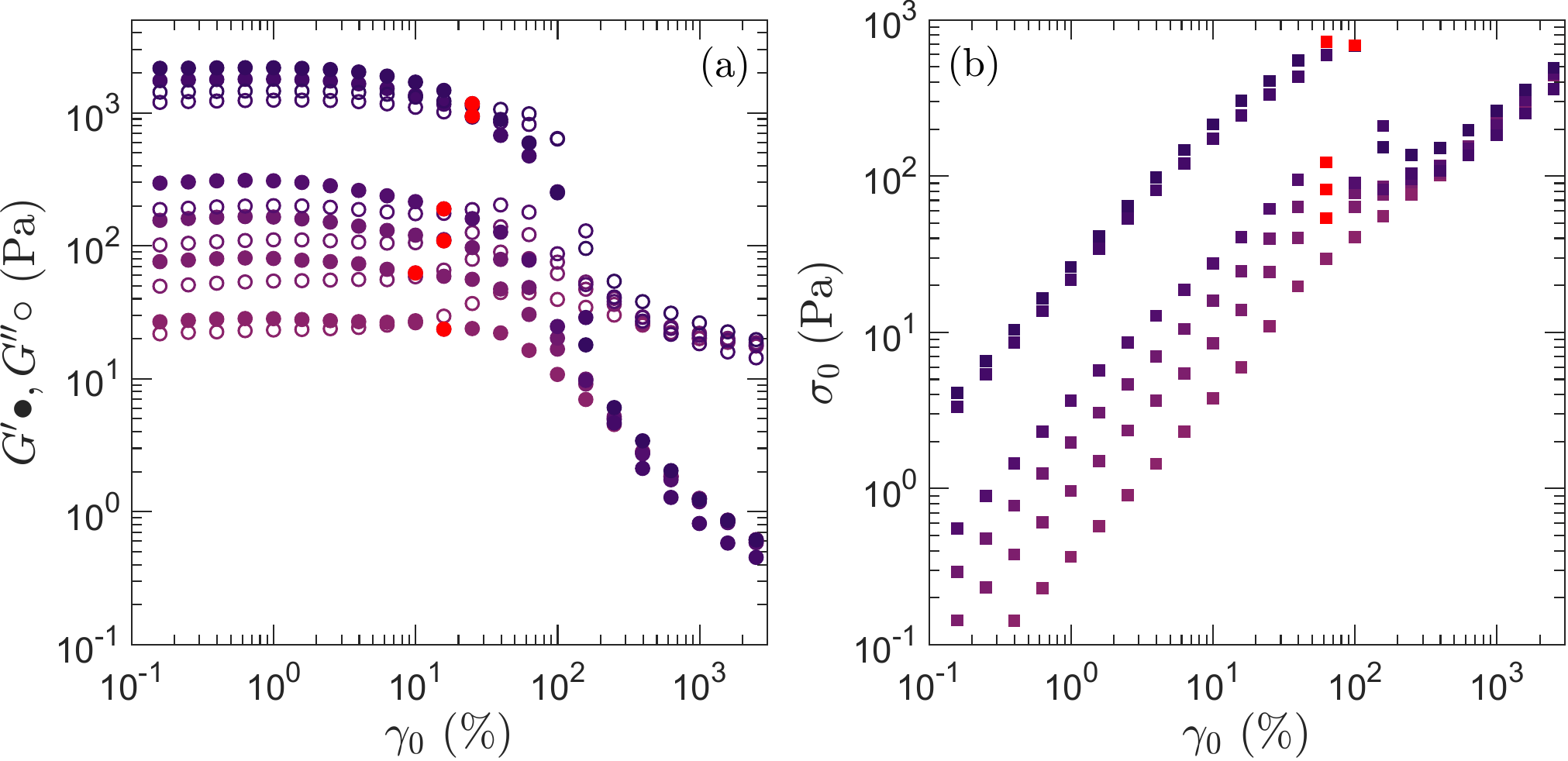}
     \caption{ Nonlinear rheology of latex suspensions at $t=446~\rm s$ and $T=24$, 25, 26, 28, 29, and $30^\circ\rm C$. (a) Apparent moduli $G'$ and $G''$ vs.~strain amplitude $\gamma_0$ at fixed frequency $\omega=2\pi~\rm rad.s^{-1}$. Red dots mark the crossover of $G'$ and $G''$, defining the yield stress $\sigma_y$. (b) Stress amplitude $\sigma_0$ vs.~$\gamma_0$ for the same measurements. Red squares mark the stress maxima $\sigma_m$. The corresponding $\sigma_y$ vs.~$G'$ and $\sigma_m$ vs.~$G'$ data are included in the insets of Fig.~4 of the main text.}
     \label{fig:FigS5} 
 \end{figure*}

\end{widetext}

\end{document}